\documentclass[aps,prl,reprint]{revtex4-1}

\usepackage[dvips]{graphicx}
\usepackage{amsmath}

\begin{document}

\title{Many-body effects in strongly-disordered III-nitride quantum wells: interplay between carrier localization and Coulomb interaction}

\author{Aurelien David}
\email{aurelien.david@polytechnique.org}
\affiliation{Soraa Inc., 6500 Kaiser Dr. Fremont CA 94555}
\author{Nathan G. Young}
\affiliation{Soraa Inc., 6500 Kaiser Dr. Fremont CA 94555}
\author{Michael D. Craven}
\affiliation{Soraa Inc., 6500 Kaiser Dr. Fremont CA 94555}

\date{\today}

\begin{abstract}
The joint impact of Anderson localization and many-body interaction is observed in the optical properties of strongly-disordered III-nitride quantum wells, a system where the Coulomb interaction and the fluctuating potential are pronounced effects with similar magnitude. A numerical method is introduced to solve the 6-dimensional coupled Schrodinger equation in the presence of disorder and Coulomb interaction, a challenging numerical task. It accurately reproduces the measured absorption and luminescence dynamics of InGaN quantum wells  at room-temperature: absorption spectra reveal the existence of a broadened excitonic peak, and carrier lifetime measurements show that luminescence departs from a conventional bimolecular behavior. These results reveal that luminescence is governed by the interplay between localization and Coulomb interaction, and provide practical insight in the physics of modern light-emitting diodes.
\end{abstract}
\pacs{}

\maketitle

The physics of compound semiconductors is influenced by many-body interaction as well as disorder. In conventional semiconductors, disorder effects are weak compared to the Coulomb interaction, and disorder can be treated as a perturbation to the excitonic center-of-mass. This has led to multiple signatures at cryogenic temperatures in the widely-studied system of GaAs/AlGaAs quantum wells (QWs) with disordered interfaces, using absorption and luminescence as probes for the quantum state of the system \cite{Weisbuch81,Wang90,Zrenner94,Brunner94,Gammon96,Savona00,Intonti01,Langbein02,Yayon02}.

More recently, the mastery of III-nitride compounds --the material system of choice for modern light-emitting diodes (LEDs)-- has renewed interest in the physics of luminescence of disordered semiconductors by giving access to a new physical regime. Indeed, these materials constitute a remarkable system where the two effects are strongly pronounced: Coulomb interaction leads to an excitonic binding energy of tens of meVs, with an associated Bohr radius of a few nm; meanwhile, Anderson localization of carriers stems from inherent random alloy disorder, with fluctuations in potential energy of $\sim 100$ meV and typical dimensions of a few nm. Therefore, both phenomena occur with a similar scale of energy (large enough to be relevant at room temperature) and distance, and must be considered on equal footing rather than perturbatively. III-nitride QWs thus offer an ideal testbed to study the unique regime of strong disorder with strong Coulomb interaction, with the practical perspective of better understanding the fundamental limits of this material for LED applications.

Recent experimental work has shown evidence of localization effects in III-nitride QWs \cite{Chichibu06,Piccardo17,Hahn18,Blenkhorn18} with signatures reminiscent of that in conventional semiconductors \cite{John86,Wang90,Zrenner94,Brunner94,Gammon96,Savona00}; on the other hand, direct observation of Coulomb-interaction effects has been elusive. From a theoretical standpoint, alloy disorder \cite{Watson-Parris11,Schulz15,Aufdermaur16,Li17,Jones17} and many-body effects \cite{Chow99,Piprek_Book07_ch7} have been investigated independently. Their joint consideration is a more complex task; it was first tackled recently in Refs.~\cite{Schulz15,Tanner18}, with the limitation that the Coulomb-interacting problem is solved perturbatively over a basis of a few free-carrier eigenstates. Overall, these theoretical investigations only predict moderate corrections to disorder-free models, and have not been validated experimentally.

In this letter, we present direct experimental evidence of the effects of localization and many-body interaction on the room-temperature optical properties of high-quality InGaN QWs, and show that their magnitude well exceeds the aforementioned predictions. We introduce an advanced numerical model which treats these on equal footing by solving the full 6-dimensional Schrodinger equation. This powerful approach quantitatively reproduces experimental results, confirming that the interplay between the two effects is essential in understanding III-nitride luminescence and the resulting LED efficiency.

\begin{figure*}[!!!thhhhhhhhhhhhb]
\includegraphics[width=\textwidth]{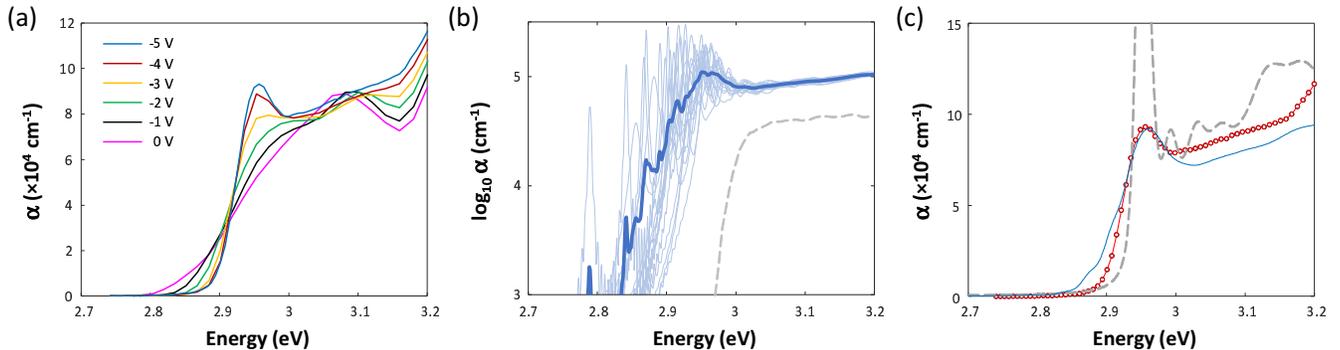}
\caption{Absorption properties of a 3.5nm-thick InGaN QW. (a) $\alpha$ measured by photocurrent at various reverse biases, showing the progressive appearance of an excitonic resonance. (b) Modeled $\alpha$ (in log scale) in flat-band conditions. Thin lines: individual configurations. Thick line: average of 40 calculations. Dashed line: average without Coulomb interaction, for comparison. (c) Comparison of measured (solid line with dots) and modeled (solid line) $\alpha$, showing excellent agreement. Dashed line: modeled $\alpha$ without disorder.}
\label{Fig1}
\end{figure*}

We begin by investigating the absorption coefficient ($\alpha$) in InGaN QWs, since excitonic effects are commonly manifested by sharp absorption peaks. Such peaks are indeed observed in bulk GaN samples at cryogenic temperature \cite{Dingle71}. In contrast, measurements on QWs usually result in broad spectra which lack excitonic signatures. We illustrate this using a sample, grown on a bulk GaN substrate by MOCVD, having a 3.5 nm-thick InGaN QW with [In]$=13\%$ placed at the center of a p-i-n junction. Its room-temperature absorption, measured by photocurrent spectroscopy \cite{Collins86,Piccardo17}, lacks distinct features, as shown in Fig.~\ref{Fig1}(a).

Such featureless absorption is attributed to the inhomogeneous broadening caused by the strong polarization field present across the QW, and to the reduction in Coulomb interaction due to electron-hole wavefunction separation. To remove these effects, we perform photocurrent measurements under reverse bias to compensate for the polarization field. As the bias is varied, the absorption spectrum progressively sharpens until the QW reaches flat-band conditions (Fig.~\ref{Fig1}(a)), where an excitonic peak is observed \cite{David10b}. We have systematically observed similar spectra in other samples of varying QW design (data not shown). This excitonic peak constitutes a rare direct manifestation of many-body effects at room temperature in InGaN QWs, and offers an opportunity to validate theoretical models.

We now present a numerical model which predicts such optical features by taking into account alloy disorder and Coulomb interaction. Since both effects are of similar magnitude, they cannot be treated perturbatively as has been done in other material systems. We place ourselves in the two-band effective mass approximation; this is appropriate to describe the near-band-edge optical behavior. The electron-hole Hamiltonian reads:

\begin{equation}
\label{Eq:H}
H_r = -\frac{\hbar^2}{2m_e} \Delta_e + V_e - \frac{\hbar^2}{2m_h} \Delta_h + V_h +V_{C}
\end{equation}

Here $\Delta_{e,h}$ are the Laplacian operators on the electron and hole coordinates, $V_{e,h}$ the electrostatic potentials for the electron and hole (including alloy disorder), and $V_{C}$ the Coulomb interaction term.

Because it includes disorder and Coulomb interaction, Eq.~(\ref{Eq:H}) is a 6-dimensional problem (3 dimensions per carrier, coupled together), making it very challenging numerically. The standard approach (summing optical transitions over eigenstates of $H_r$) is prohibitive as several thousand eigenstates are required for a proper description of the optical joint density of states (JDOS). Instead, we proceed by solution of the time-dependent Schrodinger equation in real space, following the general approach of Refs.~\cite{Glutsch96,Glutsch04}. In short, the optical polarization is propagated in time, which directly yields the JDOS and absorption without requiring eigenstates. This makes the method computationally-efficient, although it has not previously been applied to problems with such complexity.  We mention important aspects of the model below; implementation details are in the Supplemental Material (SM).

The Hamiltonian is discretized in real space with a standard finite difference scheme. To account for alloy disorder in $V_{e,h}$, we create smoothed numerical maps of atomic potentials following the approach of Refs.~\cite{Watson-Parris11,Piccardo17}.

The Coulomb term $V_{C}$ deserves caution. Since $V_C \sim 1/r$, a naive discretization diverges at $r=0$, causing well-known numerical difficulties. The often-proposed simple regularization $V_C \sim 1/(r^2+a^2)^{1/2}$ (with $a$ an empirical short-scale constant) suffers from poor convergence. Instead, we have investigated two more sophisticated approaches: the so-called ground-state scheme \cite{Glutsch96,Glutsch04} and asymptotic-behavior-correspondence (ABC) scheme \cite{Gordon06}. Both lead to near-identical results in our simulations. As we will see hereafter, the model will ultimately need to account for carrier-screening effects. Therefore, we select the ABC scheme because it can straightforwardly be generalized to arbitrary carrier populations \footnote{In contrast, the ground-state scheme requires a preliminary calculation of the ground state for the 1-dimensional excitonic problem, which only exists below the Mott density}. The Coulomb term thus reads: $V_{C}=-e^2/4\pi \epsilon \bar{r}$, with $\bar{r}$ the effective radius derived from the ABC scheme and $\epsilon$ the dielectric constant of GaN.

With this model, we compute the optical absorption of a QW in flat-band conditions. Fig.~\ref{Fig1}(b) shows several calculations with different configurations of the alloy distribution, and the average of 40 calculations. For each configuration, sharp excitonic peaks are observed. Near the band edge these peaks are dense; each configuration also displays a few deeply-localized excitonic peaks: these correspond to excitonic states stemming from Anderson-localized holes at random In-rich locations in the QW. These deep states produce an Urbach tail for the average absorption, with a characteristic energy $\sim20$~meV, similar to the measurements of Ref.~\cite{Piccardo17}. Note that ignoring Coulomb interaction would lead to a narrower Urbach tail (7 meV, Fig.~\ref{Fig1}(b)). The Urbach tail is often considered as a manifestation of wavefunction localization\cite{John86,Schubert86}; our results show that Coulomb interaction further affects its behavior. We infer that the observed increase in Urbach energy is caused by a more pronounced Coulomb interaction among deeply-localized states.

\begin{figure*}[!!!tttth]
\includegraphics[width=\textwidth]{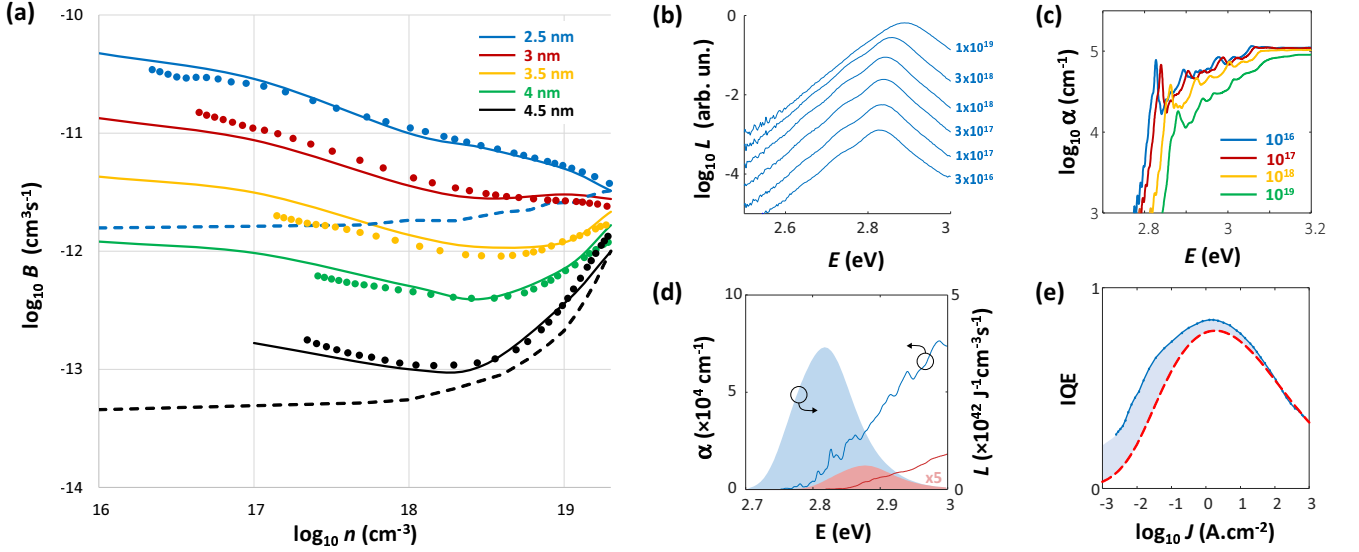}
\caption{Photoluminescence properties of InGaN QWs. (a) Effective radiative coefficient versus carrier density. Dots: experimental measurements, revealing a departure from bimolecular recombination. Dashed lines: model including disorder but ignoring Coulomb interaction, resulting in a bimolecular rate, i.e. a plateau at low density with an increase due to field-screening at high density (shown only for the thinnest and thickest QWs). Solid lines: full model. (b) Experimental luminescence spectra $L$ for a 3~nm QW, indicative of thermalized carrier distributions at all carrier densities (densities as labeled). (c) Modeled carrier-density-dependent absorption for one configuration of a 3 nm QW (densities as labeled). (d) Modeled absorption $\alpha$ (lines) and luminescence $L$ (shaded shapes) for a 3 nm QW; blue/red: with/without Coulomb interaction. (e) IQE of a 2.5 nm QW versus current density $J$. Solid line: measured IQE; dashed line: predicted IQE in the absence of Coulomb enhancement.}
\label{Fig2}
\end{figure*}

Fig.~\ref{Fig1}(c) compares the average calculated absorption (smoothed with a conservative linewidth of $15$ meV \cite{Holmes14,Hahn18}, a value low-enough to smooth numerical noise without dominating the inhomogeneous broadening) to the experimental data. An excellent agreement is obtained without adjusting any model parameter. The amplitude and shape of the excitonic peak and of the low-energy edge are well-reproduced. For comparison, Fig.~\ref{Fig1}(c) also shows a calculation where alloy disorder was ignored: the corresponding excitonic peak is much too sharp. This confirms that Anderson localization dominates the inhomogeneous broadening in flat-band absorption spectra.

Having addressed optical absorption, we turn to the more complex and important study of luminescence--more specifically luminescence dynamics, for which many-body effects can cause a departure from the conventional bimolecular radiative rate $G_r=Bn^2$ (with $B$ the radiative coefficient and $n$ the density of electrons and holes).

We study a series of QW samples similar to the previous sample, with varying QW thickness. The epitaxial structure ensures the absence of artifacts from modulation doping \cite{Langer13} and carrier escape \cite{David16a}, thus enabling a proper measurement of carrier dynamics using an all-optical differential lifetime measurement \cite{David17a}. In contrast to conventional large-signal measurements, this technique gives direct access to the lifetime and carrier density $n$ over a wide range of excitation levels, without requiring assumptions on a recombination model. By combining this with a measurement of the sample's absolute internal quantum efficiency (IQE), we obtain the effective radiative coefficient $B(n)=G_r/n^2$ \cite{David10a}. 

We measure $B$ for a series of five samples spanning QW thicknesses between 2.5 nm and 4.5 nm, as shown in Fig.~\ref{Fig2}. This data shows an intricate behavior. At high carrier density, $B$ increases for most samples; as discussed in our previous work, this is simply due to screening of the polarization field by injected carriers \cite{David17a}. Qualitatively, screening should be most pronounced for thicker samples (where the potential drop due to the polarization field is larger), as we do observe experimentally. 

The behavior at low density, however, defies common expectation: instead of a plateau (characteristic of standard bimolecular recombination), $B$ shows a clear carrier dependence, increasing up to tenfold at low density--this is most pronounced in thin QWs. The remainder of this Letter is dedicated to further investigating this remarkable trend.

One may first wonder if carrier localization can alone cause the observed increase of $B$. We have verified that this is not the case by computing the radiative rate for free carriers (i.e. carriers without Coulomb interaction) in a QW with alloy disorder. As shown as dashed lines in Fig.~\ref{Fig2}(a), a bimolecular rate is still predicted. This conclusion  is unsurprising, as the bimolecular behavior is generally robust against the details of transition selection rules \cite{Lasher64}. Furthermore, our calculations with disorder alone only lead to a small correction to $B$ from the disorder-free case (see SM): this is in line with other theoretical investigation \cite{Schulz15,Aufdermaur16,Li17,Jones17}, but does not reproduce the experimental order-of-magnitude increase in $B$, showing that disorder alone is insufficient to explain the data.

Instead, one must again consider many-body effects to account for the radiative dynamics. Studies in various material systems have shown how Coulomb interaction increases the radiative rate at low density \cite{Schlangenotto74,Chow99,Chatterjee04,Koch06,Altermatt06}. 

We therefore extend the model to encompass luminescence. Highly-accurate approaches exist for this task \cite{Kira98}; however, these are prohibitively complex for the present 6-dimensional problem. Instead, as will be detailed hereafter, we proceed at the lowest order by deriving a carrier-dependent absorption spectrum, then transforming it into a luminescence spectrum (see SM for further discussion).

The impact of carriers on absorption is included by computing a statically-screened Coulomb potential: $V_{Cs} = V_{C} \exp{(-\kappa \bar{r})}$, with $\kappa$ the Thomas-Fermi screening length (see details in SM) \cite{Haug03book}. 
This screened potential is used to compute the screened absorption spectrum, using the same procedure as before. Fig.~\ref{Fig2}(c) illustrates results for a 3 nm QW. The Coulomb enhancement is maximal at low density, and is progressively screened until the free-carrier limit is reached at high density. Note that the enhancement is comprised of excitonic peaks and the Sommerfeld factor \cite{Haug03book}, although the distinction between bound states and continuum is not well-defined in the presence of disorder.

To then obtain luminescence spectra, we make assumptions on the carrier populations. First, we assume that all the carriers exist as an electrons-hole plasma, with no exciton population. This is well-justified at room temperature considering the Saha equation \cite{Snoke08}\footnote{Specifically, assuming an exciton binding energy of 20meV, the Saha equation predicts less than 5\% of carriers in an excitonic phase  at any carrier density.}. Thus in this model, the luminescence enhancement is not due to the presence of an excitonic population, but solely to an increase of the JDOS by the Coulomb interaction--a distinction discussed in detail in Refs.~\cite{Chatterjee04,Koch06}. Second, we assume that the carriers are in quasi-thermal equilibrium, and described by Fermi-Dirac populations. This assumption may not be obvious for holes in the presence of Anderson localization; however the following considerations justify it: (i) our free-carrier computations confirm that only the lowest-energy hole states are localized, while higher-energy hole states (50meV and above) extend laterally and enable population thermalization \footnote{It can be calculated that about $25\%$ of the populated hole states are fully delocalized. These states are expected to undergo scattering with localized states and ensure thermalization.}; and (ii) experimental room-temperature luminescence spectra indeed display a thermalized tail, independent of carrier density (Fig.~\ref{Fig2}(b)).

Under these assumptions, the densities of states for free electrons and holes are derived by computing eigenstates of the respective 3-dimensional Poisson-Schrodinger equations. Several thousand states are computed, enough to generate accurate densities of states and the corresponding carrier distributions $f_{e,h}$.

Absorption and luminescence are related by the Kubo-Martin-Schwinger (KMS) relationship \cite{vanRoosbroeck54,Kubo57,Martin59,Bhattacharya12}. Here we use a generalization of this relationship, valid in disordered systems (see SM):

\begin{equation}
\label{Eq:KMS2}
L(E)=\frac{E^2n^2}{\pi^2c^2\hbar^3} \alpha \left< f_e(1-f_h) \right>
\end{equation}

where $L$ is the luminescence spectrum, $E$ the emission energy, $n$ the refractive index, $\alpha$ the absorption with Coulomb enhancement
, and $\left< \cdot \right>$ denotes an average over all pairs of electron-hole states with transition energy $E$, weighed by their wavefunction overlap. In the non-Coulomb-interacting case, Eq.~(\ref{Eq:KMS2}) exactly predicts $L$. To compute luminescence with many-body effects, we apply this relationship with carrier distributions given by their free densities of state, but replacing the free-carrier absorption with the Coulomb-interacting one \cite{Chow99}. This approximation is accurate at high enough temperature \cite{Chatterjee04}. The resulting Coulomb-enhancement of $L$ is illustrated in Fig.~\ref{Fig2}(d).

Finally, we obtain the radiative coefficient with many-body enhancement as $B=\int{L(E)dE}/n^2$. The resulting $B$ coefficients are shown on Fig.~\ref{Fig2}(a). To best match the experimental data, the modeled QW thickness was adjusted by $+0.5$ nm for all samples (a correction within the experimental uncertainty on the QW thickness). The resulting predictions are in excellent agreement with the experimental data, reproducing the relative variation of $B$ with QW thickness and its carrier-dependence.

Therefore, the behavior of the radiative rate at low current can be understood as follows. By increasing $\alpha$, many-body interaction leads to an enhanced radiative rate; this interaction is most pronounced for thin QWs (where the electron and hole wavefunctions are closely confined) and weaker for thick QWs (where the wavefunctions are separated). The enhancement is modulated by in-plane carrier localization, and is further screened by carriers, leading to a complex dependence on the structure design and carrier density. 

Importantly, this enhancement can have a profound impact on the efficiency of real-world LED devices. For instance, at a moderate carrier density $10^{17}$~cm$^{-3}$, Coulomb enhancement increases the radiative rate by an order of magnitude in thin QWs. As shown in Fig.~\ref{Fig2}(e), this causes a substantial improvement in low-current efficiency -- a regime of particular interest in applications such as micro-LEDs.

In summary, we have shown that strongly-disordered InGaN quantum wells are a remarkable system with pronounced localization and many-body effects directly observed in room-temperature optical properties. We have introduced a state-of-the-art non-perturbative model which treats alloy disorder and Coulomb interaction on equal footing, accurately and efficiently tackling this challenging numerical problem. The model confirms that the interplay between localization and Coulomb interaction dominates the optical properties of InGaN quantum emitters, making them an ideal testbed for future explorations of these complex optical effects, and shedding new insight into the efficiency of modern LEDs.


%

\clearpage

\onecolumngrid
\begin{center}
\textbf{\large Supplemental Material: \\[.5cm] Many-body effects in strongly-disordered III-nitride quantum wells: interplay between carrier localization and Coulomb interaction }
\end{center}

\twocolumngrid

\setcounter{equation}{0}
\setcounter{figure}{0}

\renewcommand{\theequation}{S\arabic{equation}}
\renewcommand{\thefigure}{S\arabic{figure}}
\renewcommand{\bibnumfmt}[1]{[S#1]}
\renewcommand{\citenumfont}[1]{S#1}

\subsection{Details of the numerical model}

We summarize the main steps of the numerical method to compute the absorption coefficient $\alpha$, as introduced in~\cite{S_Glutsch04}. $\alpha$ is related to the optical polarization $P$ (ignoring numerical prefactors):

\begin{equation}
\alpha(\omega) \propto \text{Im} \left(\int_{0}^{\infty}P(t) e^{i \omega t} \text{d}t \right)
\end{equation}

For a semiconductor structure excited at $t=0$ by a dipole $\mu \propto \delta(\textbf{r}_e-\textbf{r}_h)$, the microscopic polarization $\psi$ can be obtained by propagating the time-dependent Schrodinger equation:

\begin{equation}
i   \frac{\partial}{\partial t} \psi(\textbf{r},t) = H \psi(\textbf{r},t),
\label{Eq:TDSE}
\end{equation}

with the boundary condition $\psi(\textbf{r},t=0) = i \mu$. From this, $P$ is derived as:

\begin{equation}
P(t) = e^{-\epsilon t / \hbar} \int \mu^* \psi \text{d}\textbf{r},
\end{equation}

yielding $\alpha$. Here, $\epsilon$ is an empirical broadening energy which should be small enough that it does not dominate the inhomogeneous broadening. The key step of the calculation is the time-propagation of Eq.~(\ref{Eq:TDSE}). By discretizing $H$ and $\psi$ with a finite-difference scheme, this step takes the form of a simple matrix multiplication. The matrix is very large for a 6-dimensional problem, but efficient multiplication algorithms make the scheme tractable -- in contrast to seeking numerous eigenstates of $H$.

To construct $H$, we employ standard material parameters for the band structure of GaN~\cite{S_Vurgaftman03}. For the relative hole mass we adopt a value of 1.9, corresponding to the Luttinger parameter $A7$ far from the band edge. This choice is suited to describe hole wave packets localized in space \cite{S_Watson-Parris11}; it leads to localized hole states on a scale of a few nm, consistent with ab-initio calculations \cite{S_Schulz15,S_Aufdermaur16}. Note that the value of the hole mass is of importance to obtain proper localization results: using a lighter Gamma-point mass instead would lead to a weaker localization. We assume a band offset of $70\%$ in the conduction band; this is compatible with known values \cite{S_Moses10}, although experimental uncertainty remains \cite{S_Hurni12}. Different band offset values would slightly affect the electron localization.

\begin{figure*}[!!!thhhhhhhhhhhhb]
\includegraphics[width=\textwidth]{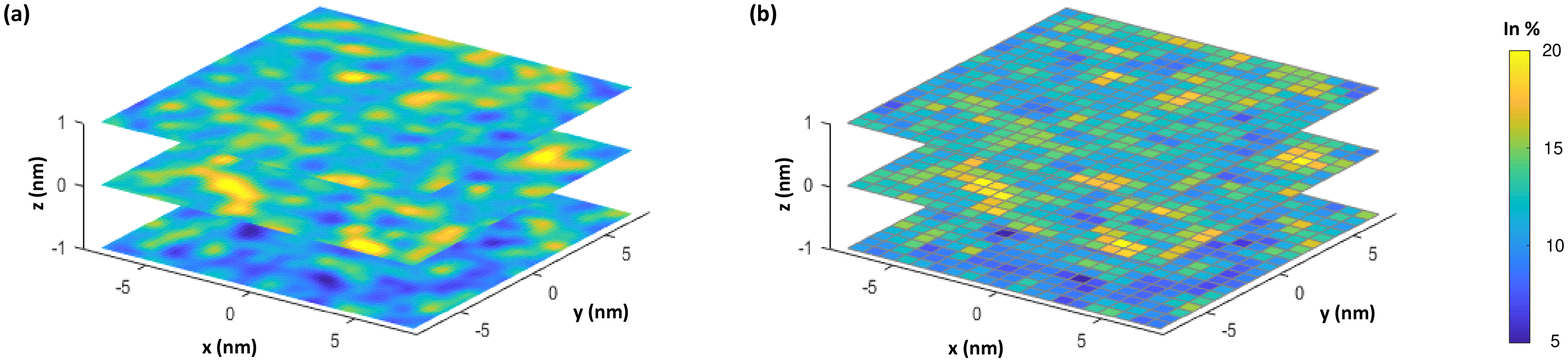}
\caption{Numerical indium distribution profiles at three planes in a 3nm QW (with average composition [In]=0.12) (a) fine grid for initial generation; (b) final calculation grid.}
\label{Suppl1}
\end{figure*}

Composition fluctuation maps are obtained following the approach of \cite{S_Watson-Parris11,S_Piccardo17}: namely, a random distribution of Ga and In atoms is generated on a fine grid, then smoothed out (by a Gaussian with 1 nm full-width at half maximum) to account for inherent averaging by carrier wavefunctions, and interpolated on the final computation grid to yield a spatial composition map (Fig.~\ref{Suppl1}). From this, the potential due to the local band gap is evaluated. The electrostatic potential is obtained conventionally by computing the strain equations and solving the resulting Poisson equation, with the p-i-n junction and applied voltage being accounted-for as boundary conditions \cite{S_Christmas05}. In this calculation, we assume that the piezoelectric constant $e_{15}$ is zero, since various works have found its effect to be small (compared to other piezoelectric terms) and uncertain, withe both positive and negative values reported \cite{S_Schulz09}. The potential contributions from the band gap and electrostatics are summed to obtain $V_{e,h}$.

The computation domain has $(xyz)$ dimensions of $16\times16\times12$~nm, with corresponding mesh steps of $0.75\times0.75\times0.5$~nm. Numerical convergence for this meshing was verified by running calculations without the Coulomb term (a 3-dimensional problem, whose lower complexity facilitates convergence testing by comparison with finer meshes). The computation domain includes several nm of GaN barrier material on either side of the QW; this is necessary due to the substantial penetration of electron wavefunctions in the barriers. To avoid spurious reflections of wavefunctions at the edges of the computation domain, absorbing boundary conditions are included \cite{S_Zambrano02}. Multiple calculations are repeated with different random alloy distributions to obtain configuration-averaged values. Computations are performed on a regular workstation; after optimization of the code for memory usage and computation speed, each calculation requires 32 Gb of memory and completes in about 2 to 10 hours, which is fast enough to perform sufficient configuration-averaging and explore the impact of the structure's parameters.

In the ABC scheme, the discretized radius operator is defined by:

\begin{widetext}
\begin{equation}
\label{Eq:abc}
\begin{split}
\frac{1}{\bar{r}} = & \frac{1}{2} \left( \sqrt{\frac{(i+1)^2}{dx^2}+\frac{j^2}{dy^2}+\frac{k^2}{dz^2}} + \sqrt{\frac{(i-1)^2}{dx^2}+\frac{j^2}{dy^2}+\frac{k^2}{dz^2}}  + \sqrt{\frac{i^2}{dx^2}+\frac{(j+1)^2}{dy^2}+\frac{k^2}{dz^2}} \right. \\
& \left. + \sqrt{\frac{i^2}{dx^2}+\frac{(j-1)^2}{dy^2}+\frac{k^2}{dz^2}} + \sqrt{\frac{i^2}{dx^2}+\frac{j^2}{dy^2}+\frac{(k+1)^2}{dz^2}} + \sqrt{\frac{i^2}{dx^2}+\frac{j^2}{dy^2}+\frac{(k-1)^2}{dz^2}}  \right) \\
& -3 \sqrt{\frac{i^2}{dx^2}+\frac{j^2}{dy^2}+\frac{k^2}{dz^2}},
\end{split}
\end{equation}
\end{widetext}

where $(dx, dy, dz)$ are the mesh steps, $i$ is defined as

\begin{equation}
\label{Eq:abc2}
i = \frac{X_e-X_h}{dx},
\end{equation}

and similar equations define $j$ and $k$.

\subsection{Luminescence with disorder}

Here, we discuss luminescence for free carriers, i.e. in the absence of Coulomb interaction. In this case, luminescence can simply be calculated by computing enough eigenstates and performing a sum over all transitions, weighed by their wavefunction overlap \cite{S_Lasher64}. This approach is valid whether alloy disorder is considered or not, and enables us to evaluate the variation in the radiative rate caused by disorder. For these calculations, we use a fine spatial mesh of 0.25~nm and a lateral domain size of 20~nm. We find that, both without and with disorder, the radiative rate is bimolecular at low carrier density (with an additinoal field-screening effect at high density). Hereafter, we consider the value of the radiative coefficient $B$ at low density.

$B$ strongly depends on the thickness of the QW, which modulates the electron-hole overlap. Therefore, when comparing QWs without and with alloy disorder, the same QW thickness should be considered. However, we are faced with an ambiguity in defining the thickness of a disordered QW, since its interfaces are not well-defined. We consider two possible definitions: (i) the thickness is the full-width at half maximum of the average indium distribution across the epitaxial direction; (ii) the thickness is defined such the same total number of In atoms is present in QWs without and with disorder.

\begin{figure}[!!!thhhhhhhhhhhhb]
\includegraphics[width=8cm]{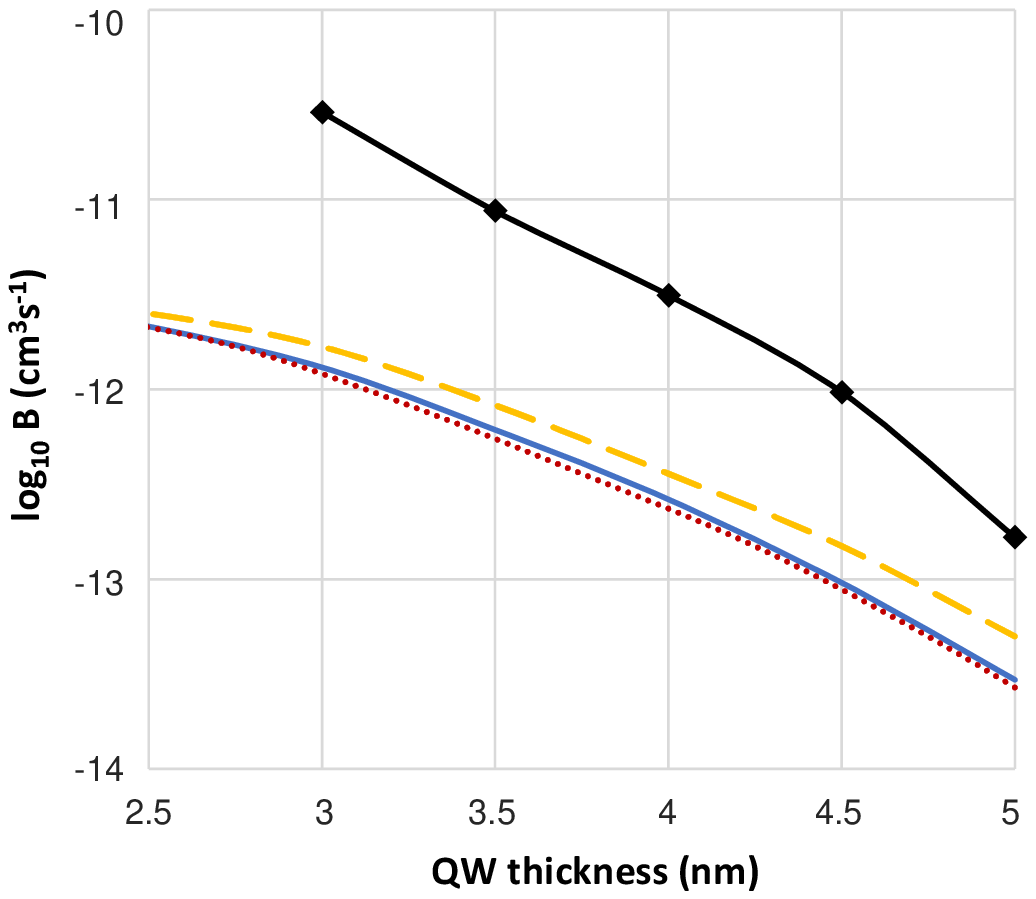}
\caption{Modeled radiative coefficient $B$ against QW thickness. Solid line: QW without disorder. Dashed/dotted lines: QW with disorder, using definitions (i) and (ii) for the QW thickness respectively. Line with symbols: QW with disorder and Coulomb interaction (using definition (i) for the QW thickness, and $n=10^{17}$~cm$^{-3}$).}
\label{Suppl2}
\end{figure}

The results of these calculations are shown in Fig.~\ref{Suppl2}. Depending on the definition for the disordered QW thickness, $B$ is either slightly smaller or larger than in a disorder-free QW. Therefore, due to this ambiguity, one cannot state whether disorder increases or decreases $B$. This ambiguity is not addressed in other modeling work, which alternately predicted a decrease \cite{S_Aufdermaur16} or an increase \cite{S_Jones17} of $B$. Regardless, the modeled effect of disorder alone is small: a few tens of \%. This stands in contrast to the model including disorder and many-body interaction (also shown on Fig.~\ref{Suppl2}), which predicts a much stronger variation with thickness and is in good agreement with our experimental data.

\subsection{Many-body luminescence}

In general, the electron-hole plasma can influence many-body interaction in various ways. First, the magnitude of the Coulomb interaction is decreased by free-carrier screening. This effect is included in our screened Coulomb potential model (see below). Second, state-filling by carriers further influences the Coulomb interaction. The interband (electron-hole) state-filling effect is accounted-for in the KMS relationship (Eq. 2). The intraband terms (electron-electron and hole-hole repulsion) are ignored; this is suitable because their two main consequences are bandgap renormalization (whose consideration is not essential to compute the luminescence rate) and an influence on carrier populations (which we assume to be in quasi-equilibrium anyway). The level of our approximations is comparable to that of \cite{S_Banyai86}.

For the screened Coulomb potential, we use the simple RPA approximation where $\kappa^2=e^2 (dn/d\mu_{e}+dp/d\mu_{h})/\epsilon$, with $n,p$ the electron and hole densities and $\mu_{e,h}$ the respective quasi-Fermi levels. More sophisticated models of the dielectric function could be considered, but would only bring second-order corrections to the present results. The derivative terms in $\kappa$ are evaluated numerically--incidentally, we find that the numerical derivatives closely match the analytical derivative formula for an ideal three-dimensional density of states (namely, $dn/d\mu_{e} = n/k_BT $) \cite{S_Haug03book}.

We now discuss the KMS relationship (also know as van Roosbroeck-Shockley relation). This relationship relates luminescence and absorption, and was originally expressed as \cite{S_vanRoosbroeck54,S_Kubo57,S_Martin59}: 

\begin{equation}
\label{Eq:kms1}
L=C \frac{\alpha}{e^{(E-E_f)/kT}-1}
\end{equation}

With $C=E^2 n^2/\pi^2 c^2 \hbar^3$ and $E_f$ the Fermi level splitting. This form is only valid at low density, before the Fermi levels cross into the bands. A more general form, also valid at high density, is \cite{S_Bhattacharya12}:

\begin{equation}
\label{Eq:kms2}
L=C \frac{\alpha (f_v-f_c)}{e^{(E-E_f)/kT}-1}=C \alpha f_c (1-f_v)
\end{equation}

Here $\alpha$ is the `bare' absorption coefficient in the absence of electrons and hole populations. The `loaded' coefficient, taking occupation into account, is $\alpha^*=\alpha (f_v-f_c)$ \cite{S_Bhattacharya12}. 

These expressions pertain to systems with in-plane symmetry, where each electron-hole pair state (with a given in-plane wavevector) corresponds to a specific transition energy. In disordered systems, this is no longer the case: several electron-hole pairs can provide transition at the same energy, so that the band occupation $f_c (1-f_v)$ can no longer be assigned to a specific energy.

To deal with this, we return to the definition of luminescence and absorption in an energy band d$E$, expressed as a sum over states (ignoring prefactors for simplicity):

\begin{eqnarray}
\label{Eq:kms3}
L(E)\text{d}E  & \propto & \sum  I^2 f_e(1-f_h),
\\
\alpha(E)\text{d}E   & \propto  & \sum  I^2 
\end{eqnarray}

Here the sum runs over all states within the energy band d$E$, and $I$ is the electron-hole overlap integral modulating the intensity of each transition. Therefore, absorption and emission remain related by a detailed-balance relationship:

\begin{equation}
L = C \alpha \frac{\sum I^2 f_e(1-f_h)}{\sum I^2} = C \alpha \left< f_c (1-f_v) \right>
\end{equation}

At low carrier density, $\left< f_c (1-f_v) \right>$ reduces to the Boltzmann limit for all electron-hole pairs and expression~(\ref{Eq:kms1}) is recovered. In the absence of disorder, all pairs with allowed transitions have the same occupation factor and expression~(\ref{Eq:kms2}) is recovered.

\begin{figure}[!!!thhhhhhhhhhhhb]
\includegraphics[width=8cm]{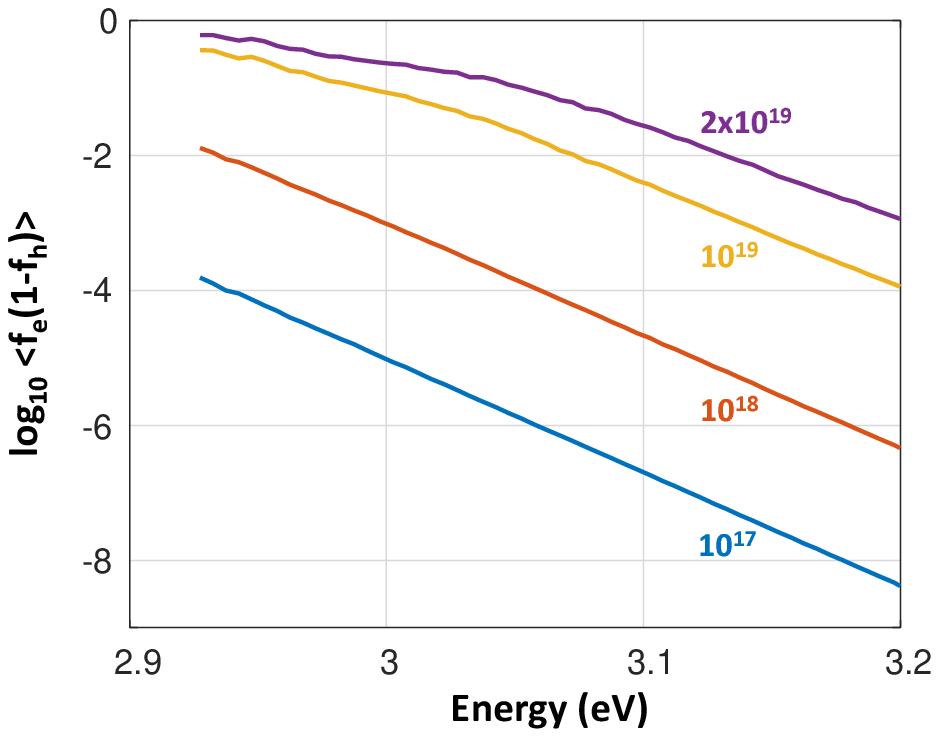}
\caption{Carrier occupation function $\left< f_e(1-f_h) \right> $ for a  3nm QW ([In]=12\%), at various carrier densities (as labeled). At low density, the usual Boltzmann limit is recovered.}
\label{Suppl3}
\end{figure}

For the many-body case, we replace the bare absorption coefficient $\alpha$ with its many-body value, obtained from Eq.~1. Note that $\left< f_c (1-f_v) \right>$  can only be calculated explicitly down to the lowest energy transition of the free electron-hole plasma, whereas the many-body absorption begins at a lower energy due to excitonic resonances. However, this occupation function displays a smooth behavior on a log scale, as shown on Fig.~\ref{Suppl3}, so that it can be extrapolated to slightly lower energy (a few tens of meV) to apply the KMS relationship.

\end{document}